\documentclass[aps,preprint,a4paper,showpacs,showkeys,superscriptaddress]{revtex4-1}
\usepackage{latexsym}
\usepackage{amsmath,amssymb}
\usepackage{graphicx}
\usepackage{subfigure}
\usepackage{xcolor}
\usepackage{hyperref}     % color hypertext for pdflatex
\hypersetup{colorlinks,%
  citecolor=blue,%
  linkcolor=cyan,%
  pdftex}

\begin{document}

%\preprint{arXiv:yymm.nnnn [gr-qc]}

\title{Something special at the event horizon}

\author{Myungseok Eune}%
\email[]{eunems@smu.ac.kr}%
\affiliation{Department of Computer System Engineering, Sangmyung
  University, Cheonan, 330-720, Republic of Korea}%

\author{Yongwan Gim}%
\email[]{yongwan89@sogang.ac.kr}%
\affiliation{Department of Physics, Sogang University, Seoul 121-742,
  Republic of Korea}%

\author{Wontae Kim}%
\email[]{wtkim@sogang.ac.kr}%
\affiliation{Department of Physics, Sogang University, Seoul 121-742,
  Republic of Korea}%

\date{\today}

\begin{abstract}
  We revisit the free-fall energy density of scalar
  fields semi-classically by employing the trace anomaly on a
  two-dimensional Schwarzschild black hole with respect to various black hole
  states in order to clarify whether something special 
  at the horizon happens or not.  For the Boulware state, the energy
  density at the horizon is always negative divergent, which is
  independent of initial free-fall positions.  However,  in the Unruh state the
  initial free-fall position is responsible for the energy
  density at the horizon  and there is a critical point to
  determine the sign of the energy density at the horizon.  In
  particular, a huge negative energy density appears when the freely falling observer   
  is dropped just near the horizon.  For the
  Hartle-Hawking state, it may also be positive or negative depending on
  the initial free-fall position, but it is always finite. Finally, we
  discuss physical consequences of these calculations.
\end{abstract}

%\pacs{04.70.Dy, 04.62.+v, 04.60.Kz } 


\maketitle
%%%%%%%%%%%%%%
%% RevTeX Style End %%
%%%%%%%%%%%%%%

%\newcommand{\lp}{\ell_P}

\section{Introduction}
\label{sec:intro}

Hawking radiation has invoked some issues which are of relevance to
not only information loss problem in
quantum gravity theory~\cite{Hawking:1974sw, Hawking:1976ra} but
also black hole complementarity~\cite{Susskind:1993if,
  Susskind:1993mu, Stephens:1993an}. The latter states that there are no
contradictory physical observations between a freely falling observer
and a distant observer since the two descriptions are complementary.
The presence of Hawking radiation indicates that the rest observer at
infinity sees the flux of particles. By the way, Unruh has argued
that ``a geodesic detector near the horizon will not see the Hawking
flux of particles"~\cite{Unruh:1980cg} and then showed that the infalling 
negative energy flux
can exist near the horizon \cite{Unruh:1977ga}. Moreover, it has been shown that
the finite flux at the horizon can be found even in the freely falling frame 
by studying the Green's function  
on the Schwarzschild black hole \cite{Candelas:1980zt}.  
Recently, it has been claimed
that a freely falling observer finds something special at the
event horizon called the firewall and burns up because of high energy
quanta~\cite{Almheiri:2012rt}, and subsequently much attention has
been paid to the firewall issue \cite{Bousso:2012as, Nomura:2012sw,
  Page:2012zc, Kim:2013fv, Giddings:2013kcj,
  Almheiri:2013hfa}.  A similar prediction referred to as an energetic
curtain has also been done based on different
assumptions~\cite{Braunstein:2013bra}.  
However, it has also been proposed that 
there is no apparent need for firewalls since the unitary evolution of black hole 
entangles a late mode located outside the event horizon with a combination 
of early radiation and black hole states, instead of either of them separately \cite{Hutchinson:2013kka},
and argued that the remaining set of nonsingular realistic states do not have 
firewalls but yet preserve information
% in Hawking radiation from black holes that form from nonsingular initial states
\cite{Page:2013mqa}.

On the other hand, there has
been some interests in radiation in freely falling frames in its own
right and it has been widely believed that the equivalence
principle tells us that the free-falling observer can not see any
radiation.  This fact is based on the classical argument of
locality but it may not be true in quantum regime
such that a freely falling observer can find quantum-mechanical
radiation and temperature~\cite{Brynjolfsson:2008uc, Greenwood:2008zg,
  Barbado:2011dx, Smerlak:2013cha, Smerlak:2013sga}. Recently, it has been claimed that the freely falling observer 
 dropped at the horizon necessarily encounters the infinite
negative energy density when the observer passes through the
horizon~\cite{Kim:2013caa}. On general grounds, one may regard this
phenomenon as a very special feature such a simplified model  \cite{Callan:1992rs} that Hawking
temperature happens to be independent of the black hole mass.
Otherwise, is there any special choice of vacuum to give rise to
the infinite energy density at the horizon?
If the existence of the infinite energy density at the
horizon turns out to be true, one may wonder what happens at the horizon 
when the frame is dropped far from 
the horizon.

Now, we would like to study the quantum-mechanical energy densities
 measured by the
freely falling observer on the two-dimensional Schwarzschild
black hole background where Hawking temperature explicitly depends
on the black hole mass.  The trace anomaly for massless scalar fields
will be employed to calculate the energy-momentum tensors along with
covariant conservation law of the energy-momentum tensors. Then,
the energy density will be characterized by three states \cite{Birrell:1982ix}; 
the Boulware,
Unruh, and Hartle-Hawking states in order to investigate what state is
relevant to the infinite energy density at the horizon.  If there
exists such a non-trivial effect at the horizon, then this fact will be tantamount to 
the failure of no drama condition which has been one of the
assumptions for black hole complementarity, and
this work will be the 
quantum field theoretic realization of non-trivial effect at the horizon \cite{Braunstein:2013bra}.

Now, in section~\ref{sec:ff.frame},
we encapsulate how to formulate the freely falling frame by solving
the geodesic equation of motion and present the explicit form of the
corresponding energy density. First, 
the simplest Boulware state will be
studied in section~\ref{boulware} where the energy density is always
negative and divergent at the horizon, which is interestingly irrespective of the
initial free-fall position $r_s$.  
In section~\ref{sec:unruh}, we
shall find much more non-trivial effects in the Unruh state such that the
observer finds positive radiation during free fall as long as
$r_s > r_0$ where $r_0$ is a point for the free-fall energy density to
vanish.  By the way, there is a critical point $r_c$ for the observer
to see only negative radiation during free fall for
$r_s <r_c$. The closer the initial free-fall position approaches the horizon, 
the larger negative energy density can be found,  
so that it can be divergent at the horizon eventually. 
There is also an intermediate region of $r_c < r_s < r_0$
for the observer to see negative radiation initially and to find
positive radiation finally at the horizon.  
Finally, we explain the reason why the infinite energy density at the horizon appears 
in our calculations.
In section~\ref{sec:HH}
for the Hartle-Hawking state, some similar behaviors to the case of
the Unruh state will be reproduced; however, the crucial difference
comes from the fact that the energy density at the horizon is
always finite.  After all, the calculation in this work
will show that the non-trivial
effect measured by the freely falling
observer in the semi-classical argument is sensitive to 
the initial free-fall position and the black hole states. 
Finally, we will discuss
physical consequences of this work in section~\ref{sec:discussion}.

 \section{Freely falling frame}
 \label{sec:ff.frame}

Let us start with the two-dimensional Schwarzschild black hole
governed by~\cite{Unruh:1980cg,Christensen:1977jc},
\begin{equation} %~\eqref{metric}
  ds^2 = -f(r)dt^2 +\frac{1}{f(r)} dr^2, \label{metric}
\end{equation}
where the metric function is given by $f(r)=1-2M/r$ and the horizon is
defined at $r_{\rm H}= 2M$. Solving the geodesic equation for the
metric~\eqref{metric}, the proper velocity of a particle can obtained
as~\cite{Ford:1993bw}
\begin{align}
  u^\mu = \left(\frac{dt}{d\tau}, \frac{dr}{d\tau}\right) =
  \left(\frac{k}{f(r)}, \pm \sqrt{k^2 - f(r)} \right), \label{u:k}
\end{align}
where \(\tau\) and \(k\) are the proper time and the constant of
integration, respectively.  The $k$ can be identified
with the energy of a particle per unit mass for \(k > 1\), 
which can be written as \(k = 1/\sqrt{1-v^2}\) with
\(v=dr/dt\) at the asymptotic infinity. 
In this case, the motion of the particle is unbounded,
so that the particle lies in the range of \(r\ge r_{\rm H}\). 
For \(0 \le k \le 1\), the motion of the particle is bounded such that there is a
maximum point \(r_{\rm max}\) where  the particle lies in the range
of \(~r_{\rm H} \le r \le r_{\rm max}\). 
We are going to consider a freely falling frame starting at 
$r_s= r_{\rm max}$ with zero velocity toward the black hole, which 
can be shown to be the latter case by identifying $k =\sqrt{f(r_s)}$ in Eq.~\eqref{u:k},
and thus the proper velocity of a
free-falling observer can be written as
\begin{equation} %Eq.~\eqref{eq:u} 
  \label{eq:u}
  u^\mu = \left(\frac{dt}{d\tau}, \frac{dr}{d\tau}\right)
  = \left(\frac{\sqrt{f(r_s)}}{f(r)}, -\sqrt{f(r_s)-f(r)}\right).
\end{equation}
If the observer starts to fall into the black hole at the
spatial infinity, then $f(r_s)=1$ while 
$f(r_s)=0$ for the observer to fall into the black hole just at the horizon. 
Then, the radial velocity with respect to the Schwarzschild time
becomes $v = -f(r)\sqrt{f(r_s) - f(r)} / \sqrt{f(r_s)}$ which
vanishes both at the initial free-fall position and
the horizon, and the maximum
speed occurs at $r= 6M r_s/(4M+r_s)$.  The
proper time from $r_s$ to $r_{\rm H}$ is also obtained as
\begin{equation}
  \label{propertime}
  \tau = 2M \frac{\sqrt{f(r_s)(1-f(r_s))} + \sin^{-1} \sqrt{f(r_s)}}{(1-f(r_s))^{3/2}},
\end{equation}
which is finite except for the case of the initial free fall at the asymptotic
infinity. 
So, it will take finite proper time to reach  the
event horizon when free fall begins at finite distance.

Now, in the light-cone coordinates defined by $\sigma^{\pm}=t \pm r^*$ through
$r^*=r+2M \ln(r/M-2)$ the proper velocity 
\eqref{eq:u} can be written as 
\begin{align}
  u^+ &=\frac{1}{\sqrt{f(r_s)}+\sqrt{f(r_s)-f(r)}}, \label{u:+} \\
  u^- &=\frac{\sqrt{f(r_s)} + \sqrt{f(r_s)-f(r)}}{f(r)}, \label{u:-}
\end{align}
where  $u^\pm=u^t
\pm u^r/f(r)$
and the energy-momentum tensors are expressed as~\cite{Christensen:1977jc}
\begin{align}
  \langle T_{\pm \pm} \rangle&= -\frac{N}{48 \pi} \left(\frac{2M
      f(r)}{r^3} + \frac{M^2}{r^4} \right) + \frac{N}{48}t_{\pm},
  \label{T:++}   \\
  \langle T_{+-} \rangle &= -\frac{N}{48\pi}
  \frac{2M}{r^3}f(r), \label{eq:Tpm} %Eq.~\eqref{eq:Tpm}.
\end{align}
where $N$ is the number of massless scalar fields and $t_{\pm}$ are
functions of integration to be determined by boundary conditions. The
general covariance is guaranteed by covariant conservation
law of the energy-momentum tensors. The two-dimensional trace
anomaly for scalar fields was employed to
get the non-trivial vacuum expectation value of the energy-momentum
tensors~\cite{Christensen:1977jc}, so that the two differential equations and one anomaly
equation determine the explicit form of the energy-momentum tensors
with two unknowns.

Now, the energy density measured by the free-falling observer can be
calculated as~\cite{pi, Ford:1993bw},
\begin{equation}
  \label{energy}
  \epsilon = \langle T_{\mu \nu} \rangle u^\mu u^\nu,
\end{equation}
by using the proper velocity and the energy-momentum tensor. 
In connection with Hawking
radiation, the fields are quantized on the classical background metric
in such a way that non-trivial radiation will appear and the energy
density~\eqref{energy} will not vanish even in the freely falling frame.
Substituting Eqs.~\eqref{u:+}, \eqref{u:-}, \eqref{T:++} and
\eqref{eq:Tpm} into~\eqref{energy}, the energy density 
can be
calculated as
\begin{align}  %Eq.~\eqref{eq:ed} 
  \label{eq:ed}
  \epsilon(r|r_s) &= -\frac{N}{48 \pi r^4 f(r)} \Bigg[ 8Mrf(r_s) +
  4M^2 \left(\frac{f(r_s)}{f(r)} - \frac{1}{2}\right)  - \pi r^4\left(\sqrt{\frac{f(r_s)}{f(r)}} -
    \sqrt{\frac{f(r_s)}{f(r)}-1}\right)^2t_+ \notag \\
&\quad -\pi r^4\left(\sqrt{\frac{f(r_s)}{f(r)}} +
    \sqrt{\frac{f(r_s)}{f(r)}-1}\right)^2t_-\Bigg],
\end{align}
and % If the observation point is coincident with the starting point of
% free fall, $i.e.,$ $r=r_s$, then the velocity of the freely falling
% frame becomes
% $(u^+(r_s),u^-(r_s))=(1/\sqrt{f(r_s)},1/\sqrt{f(r_s)})$ and the
% corresponding energy density is reduced as
it is reduced to
\begin{align}\label{eq:edr0}%Eq.~\eqref{eq:edr0} 
  \epsilon\ (r_s|r_s)= & -\frac{N}{48 \pi r_s^4 f(r_s)} [
  8Mr_sf(r_s)+2M^2 -\pi r_s^4(t_++t_-) ],
\end{align}
%%%%%%%%%%%%%Boulware state%%%%%%%%%%%%%%%%%%
at the special limit of $r=r_s$ 
where observation is done at the moment when free fall begins. 
Now, let us investigate characteristics 
for the energy density measured by the free-falling observer
for the Boulware, Unruh, and Hartle-Hawking states, respectively
in what follows.

 \section{Boulware state}
 \label{boulware}

\begin{figure}[hbt]
  \begin{center}
  \includegraphics[width=0.6\textwidth]{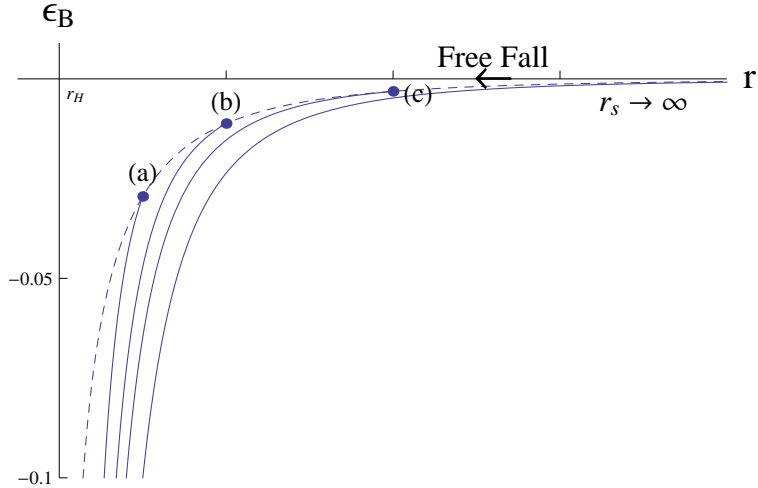}%\label{fig:}}
  \end{center}
  \caption{The energy densities in the Boulware state are plotted by
    simply choosing as $N=12, M=1$. They are
 always negative no matter where the initial free-fall positions are. 
    The solid curves represent  $\epsilon_B \ (r|r_s)$,  
    and the dotted curve does $\epsilon_B\ (r_s|r_s)$ 
     where three energy densities at $r_s$ are denoted
    by the black dots (a), (b), and (c).} 
  \label{fig:EB}  %Fig.~\ref{fig:EB}
\end{figure}

The Boulware state is obtained by choosing $t_\pm=0$, where the
energy density~\eqref{eq:ed} reads as
\begin{equation}
  \epsilon_B(r|r_s) = -\frac{NM^2}{12 \pi r^4
    f(r)} \left[\frac{2 r f(r_s)}{M} + \frac{f(r_s)}{f(r)} -
    \frac{1}{2} \right], \label{energy:B}
\end{equation}
which is always negative.  So the
freely falling observer encounters more and more negative energy
density and then eventually negative divergent one at the horizon which 
is independent of the initial free-fall position. 
If the observation is done at the moment when 
free fall begins, the energy density is reduced to 
\(\epsilon_B(r_s|r_s)=- N[ 4Mr_sf(r_s)+M^2]/[24 \pi
r_s^4 f(r_s)], \) so that the observer who starts  at the horizon
finds the divergent energy immediately 
and asymptotically vanishes without Hawking
radiation as shown in Fig.~\ref{fig:EB}.

%%%%%%%%%%%%%%%%%%%%%%%%%%%%%%
%% Fig: Boulware state %%
%%%%%%%%%%%%%%%%%%%%%%%%%%%%%%
%%%%%%%%%%%%%%%%%%%%%%%%%%%%%

 %%%%%%%%%%%%%%%%%%%%%%%%%%%%%%%%%%%%
  
%%%%%%%%%%%%%%Unruh state%%%%%%%%%%%%%%%%%%%% 

 \section{Unruh state}
 \label{sec:unruh}

The Unruh state is characterized by choosing functions of integration as
$t_+=0$ and $t_-=1/(16 \pi M^2)$ in Eq.~\eqref{eq:ed}, which yields the
energy density as
\begin{align}
  \epsilon_U(r|r_s) &= -\frac{N M^2}{12 \pi r^4 f(r)} \Bigg[ \frac{2 r
    f(r_s)}{M} + \frac{f(r_s)}{f(r)} - \frac{1}{2} - \frac{r^4}{64M^4} \left(\sqrt{\frac{f(r_s)}{f(r)}} +
    \sqrt{\frac{f(r_s)}{f(r)}-1} \right)^2 \Bigg]. \label{energy:U}
\end{align}
The energy density measured by the free-falling observer at the horizon
who falls into the black hole from $r_s$ is simplified as $ \epsilon_U (2M|r_s) =( N(63r_s^2-320M
r_s+384M^2 )/[3072 \pi M^2 r_s (r_s-2M )] $,
%%%%%%%%%%%%%%%%%%%%%%%%%%%%%%
%% Fig: Unruh state %%
%%%%%%%%%%%%%%%%%%%%%%%%%%%%%%
\begin{figure}[hbt]
  \begin{center}
  \includegraphics[width=0.6\textwidth]{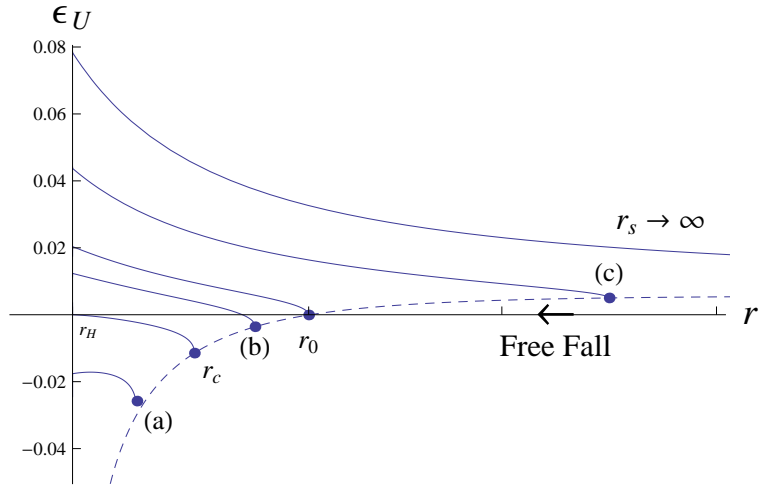}%\label{fig:}}
  \end{center}
  \caption{The energy densities for the Unruh state are plotted by
    setting $N=12,\ M=1$.  The critical point appears at $r_c \approx
    3.1M$ and the energy density vanishes at $r_0 \approx 4.2M$. 
    The solid curves are for $ \epsilon_U(r|r_s) $ and the
    dotted curve represents $\epsilon_U (r_s|r_s)$ such that
    there are largely three free-fall cases: (a) is for $r_s < r_c $, (b) is for
    $r_c <r_s<r_0$, and (c) is for $r_s  >r_0$.}
  \label{fig:EUU}  %Fig.~\ref{fig:EUU}
\end{figure}%%%%%%%%%%%%%%%%%%%%%%%%%%%%%
which is not always positive definite. 
In other words, the initial free-fall position 
is crucial to determine the sign of the energy density at the horizon in contrast to 
the Boulware case.
Specifically, the energy density at the horizon is indeed positive for
$r_s > r_0$ where $ r_0$ is the initial free-fall position for the energy
density to vanish. For instance, it is positive finite as seen 
from the case (c) in  Fig.~\ref{fig:EUU}, and it becomes 
$\epsilon_U(2M| \infty) = 21N/(1024 \pi
M^2)$ where the free-fall frame is dropped at the
spatial infinity. 
On the other hand, there is a critical point $r_c =
8(20M+\sqrt{22}M)/63$ which is defined by the point where the observer
finds the zero energy at the horizon so that the observer will
see the positive energy at the horizon as long as $r_s >r_c$.
 For $r_c < r_s
<r_0$, there appears a transition from the
negative energy density to the positive energy density, which can be seen from the
case (b) in Fig.~\ref{fig:EUU}. 
For $r_s <r_c$, the observer will see only negative radiation at the
horizon like the case (a). 
When the initial free-fall position approaches the horizon closer, 
the larger negative energy density appears.

Using Eq.~\eqref{eq:edr0}, the energy density can be obtained 
at the moment when the
free fall just begins, then the corresponding energy density is
given as \( \epsilon_U(r_s|r_s) = -NM^2 [ 4 r_s f(r_s)/M + 1 -
r_s^4/(32 M^4) ]/[24 \pi r_s^4 f(r_s)] \) which is described by the
dotted curve in Fig.~\ref{fig:EUU}.  Note that the energy density 
at the horizon $\epsilon_{U}(2M| 2M) $ is
negative divergent whereas it is positive finite
$\epsilon_{U}(\infty|\infty) \rightarrow \pi (N/12)T_{\rm H}^2$ at the
asymptotic infinity, where $T_{\rm H}$ is the Hawking temperature.
 %%%subFig%%%%%%%%%%%%%%%%%%%%%%%%%%%%%%%%%%%%%%%%%%%%%%%%%%%%%%%%%%%%%%%%%%%%%
%\begin{figure}[t]
%\begin{center}
%\subfigure[{$\epsilon_U(2M,r_s)$}]{\includegraphics[width=0.75\textwidth]{EU2}}
%\subfigure[{$\epsilon_U(r_s,r_s)$}]{\includegraphics[width=0.75\textwidth]{EU3}}
%\end{center}
%\caption{Unruh state. $N=12, M=1$, $r_{c1} \approx 3.1M, r_{c2} \approx 4.2M$. Note that $r_s$ is the starting point of freely falling observer. Whether observer meets negative energy or not is determined by where he start freely falling from. You can see it in Fig.~\ref{fig:EU}.} 
%\label{fig:EU2Mr0}%Fig.~\ref{fig:EU2Mr0}
%\end{figure}
%%%%%%%%%%%%%%%%%%%%%%%%%%%%%%%%%%%%%%%%%%%%%%%%%%%%%%%

In particular, we would like to explain why the freely falling observers who are
moving slowly with respect to the black hole when they pass through the horizon should see very high
(negative) energy density. Actually, the conventional wisdom is that the freely falling observer near the
horizon cannot see any outgoing Hawking radiation as $\langle T_{--}\rangle=0$. 
It can be easily understood from the Unruh effect \cite{Unruh:1980cg}
which tells us that the frame near the horizon can be described as the local-flat metric in terms of 
the Kruskal coordinates on the
Schwarzschild black hole, so that the corresponding observer can be regarded as the freely falling observer
while the fiducial observer sees radiation because the observer is now on the accelerated frame. 
Moreover, in the collapsing black hole described by the Unruh state, it was shown that
the energy flow across the future horizon is seen to be negative, $\langle T_{++} \rangle <0 $,
 since the corresponding 
positive energy would flow out to infinity \cite{Unruh:1977ga}. All these facts can also be confirmed 
by using Eqs.~\eqref{T:++} and \eqref{eq:Tpm}. 

In this work, we employed the energy density which consists of three components
of $\langle T_{++} \rangle,~\langle T_{--} \rangle,~\langle T_{+-} \rangle$, 
while the energy-momentum tensors related to the fluxes have been discussed in the previous works. 
Explicitly, the energy density~\eqref{energy}
can be reduced to $\epsilon = \langle T_{++} \rangle u^+u^+ $
at the horizon since $ \langle T_{--}\rangle=\langle T_{+-}\rangle=0$ there.
Note that it does not vanish but also is negative because of non-vanishing ingoing negative flux as 
$\langle T_{++} \rangle =-N/(768 \pi M^2) <0 $ from Eq.~\eqref{T:++}. At the horizon, the non-vanishing energy density 
is related to the non-vanishing ingoing energy momentum tensor as it should be.
To explain the reason why the high energy density appears near the horizon for a
very slowly falling frame,
let us rewrite the free-fall energy density \eqref{energy}
 as $\epsilon = \langle T_{tt} \rangle u^tu^t$ 
in the normal coordinates
where the radial velocity is 
fixed as $u^{r}=0$ for convenience when the observer is dropped from rest at $r=r_s$.
Note that the time component of the velocity at the stating point of $r_s$ by definition 
becomes $u^t=dt/d\tau=1/\sqrt{f(r_s)}$ which
is larger than one because the function $f(r_s) $ is less than one near the horizon, so that
$dt > d\tau$ where 
$dt$ is a time measured by the fiducial observer 
 and $d\tau$ is a proper time measured by the freely falling observer. 
Moreover, it shows that the gravitational time dilation effect 
is much more significant when the observer is dropped close to the horizon.  
By the way, as a corollary to this fact, the frequency in the freely falling frame is higher than
that in the fixed frame, so that this factor contributes to the energy density. 
Therefore, it eventually becomes the high energy density of $\epsilon(r_s|r_s)  =-N/(768 \pi M^2 f(r_s))$
near the horizon,
where $r_s$ represents the starting position when the observer is dropped from rest.
On the other hand, if the freely falling observer starts with the non-zero initial velocity at 
a certain point from the horizon, then the observer can see the positive energy at that instant because 
$\langle T_{tr} \rangle$ with $u^r \neq 0$ gives rise to the positive contribution to the energy density.

One more thing to be mentioned in this section is that we could calculate
 the free-fall energy density not only at any finite distance but also near the horizon and at infinity 
in the simplified context which is one of  the advantages of the two-dimensional model, 
so that we could further discuss
the critical point to characterize the  positive energy zone and the negative energy zone 
by solving the exact geodesic equation analytically.  
The result shown in Fig.~\ref{fig:EUU} is physically compatible with the previous one that 
the positive energy flux would flow out to infinity while a corresponding amount of negative
energy flux would flow down the black hole \cite{Unruh:1977ga}, 
so that the area of horizon decreases at a rate
expected positive energy flux at infinity \cite{Candelas:1980zt}. 

 %%%%%%%%%%%%%%Hartle-Hawking state%%%%%%%%%%%%%%
 \section{Hartle-Hawking state}
 \label{sec:HH}

\begin{figure}[hbt]
  \begin{center}
  \includegraphics[width=0.6\textwidth]{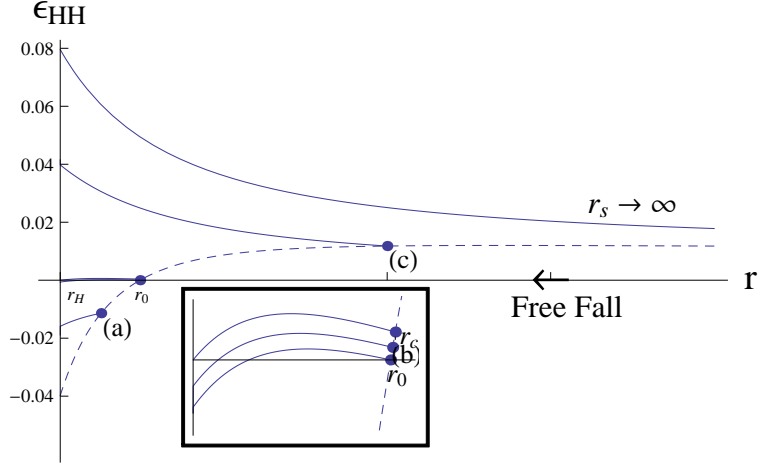}%\label{fig:}}
  \end{center}
  \caption{The energy densities in the Hartle-Hawking state are plotted
    by setting $N=12,\ M=1$. The sold curves describe $\epsilon_{HH} (r|r_s)$
    and the dotted curve represents $\epsilon_{HH} (r_s|r_s)$. 
    There are largely three free-fall cases (a) is for $r_s < r_0$, (b)
    in the box is for $r_0< r_s <r_c$, and (c) is for $r_s >
    r_c$, where $r_0 \approx 2.98M$  and $r_c=3M$.}
  \label{fig:EHH}  %Fig.~\ref{fig:EHH}
\end{figure}

For the Hartle-Hawking state, let us take 
$t_\pm=1/(16\pi M^2)$ in Eq.~\eqref{eq:ed}, then the energy density
can be obtained as
\begin{align}
  \label{hh}
  \epsilon_{HH}(r|r_s) = & -\frac{N M^2}{12 \pi r^4 f(r)}
  \left[\frac{2 r f(r_s)}{M}  - \left(\frac{r^4}{16M^4} - 1\right)
    \left(\frac{f(r_s)}{f(r)} - \frac{1}{2}\right)\right]. 
\end{align}
The free-falling observer at $r_s$ toward the black hole will find the finite
energy density at the horizon of \(\epsilon_{HH}(2M|r_s)=N (r_s - 3M) /
(48\pi M^2 r_s)\). In particular, it becomes
$\epsilon_{HH}(2M|\infty) =
N/(48M^2\pi)$ when the observer is dropped at
spacial infinity at rest. 
There is the point $r_0$ where the energy density measured 
in the free-falling frame vanishes; however,
the crucial difference from the Unruh case is that the freely falling
observer starting at $ r_s >r_0$ may encounter alternatively the positive energy and
the negative energy density during the free fall as shown in the
case (b) in Fig.~\ref{fig:EHH},
whereas only 
the positive energy density appears 
in the Unruh state. 
%%%%%%%%%%%%%%%%%%%%%%%%%%%%%%
%% Fig: Hartle-Hawking state %%
%%%%%%%%%%%%%%%%%%%%%%%%%%%%%%
%%%%%%%%%%%%%%%%%%%%%%%%%%%%%
 There is also the critical point $r_c$ to
characterize the sign of the energy density at the horizon as has been
done in the Unruh case, so that the freely falling observer at the horizon will see the
positive energy density for $r_s >r_c$ and the negative
energy density for $r_s <r_c$ . Moreover,  the
observer will necessarily find a transition from the positive
energy density to the negative energy density for $r_0<r_s <r_c$.
Note that $r_0$ and $r_c$ in the Hartle-Hawking state are smaller than
those in the Unruh state, respectively, which is shown in
Fig.~\ref{fig:EHH}.

 At $r =r_s$, the energy density in the Hartle-Hawking state 
from Eq.~\eqref{eq:edr0} becomes \(
\epsilon_{HH}(r_s|r_s) = - N [8Mr_sf(r_s) + 2M^2 - r_s^4/(8 M^2) ] /
[48 \pi r_s^4 f(r_s)], \) where the behavior of the energy density is 
described by the dotted
curve in Fig.~\ref{fig:EHH}.  
Explicitly, when $r_s = 2M$, it becomes
negative finite as $\epsilon_{HH}(2M|2M) \rightarrow -N/(96\pi M^2)$
which is contrast to the infinite energy density in the Unruh state.
At the asymptotic infinity, it is finite $
\epsilon_{HH}(\infty|\infty) \to \pi N T_{\rm H}^2/6$, and the energy
density in the Hartle-Hawking state is two times that of the Unruh
state, $i.e.$, $\epsilon_{HH}(\infty|\infty)=2\epsilon_U(\infty|\infty)$.

 \section{Discussion}
 \label{sec:discussion}
At first sight, it seemed that the freely falling observer could not find
any radiation based on the intuitive argument that the surface gravity translated into 
the radiation temperature vanishes in local inertial frames; 
however, the explicit calculation shows that
radiation can exist even in freely falling frames and moreover 
it depends on the free-fall position unequivocally for certain black hole states.
For the
Boulware state, the energy density is always negative divergent at the
horizon, which is independent of the initial free-fall positions. 
For the Hartle-Hawking state, the energy density at the horizon is negative finite 
when free fall toward the black hole begins at $r_s < r_c$ while it is
positive finite at the horizon for $r_s > r_c$, where $r_c=3M$ is the
critical initial free-fall position to determine the sign of the energy density
at the horizon.  In particular, the Unruh state describing  
the collapsing black hole gives a slightly larger critical
value of $r_c \approx 3.1 M$ compared to that of the Hartle-Hawking
state. As expected, the energy density at the horizon in the Unruh state behaves as mixed between
the Boulware state and the Hartle-Hawking state in that the energy density at the horizon 
can be either divergent or finite.  Therefore, the energy density
measured by the freely falling observer in the semi-classical black
hole is characterized by mainly two conditions of both the initial free-fall
position and the black hole state.

In connection with black hole complementarity, 
let us drop the particle at a finite distance near the
horizon such that it will take finite proper time for the freely falling observer to reach the horizon. 
Subsequently, the particle will reach
the origin without any drama at the horizon.  On the other hand, the
distant observer using the Schwarzschild coordinates feels that the
particle is still at rest at the horizon so that the observer cannot
see the particle crossing the horizon forever. But, the particle will
appear awkwardly through Hawking radiation eventually.  To resolve
this problem, the notions of the membrane paradigm and the
stretched horizon with black hole complementarity are helpful in
understanding the relation between the infalling particle and Hawking
radiation \cite{Susskind:1993if,
  Susskind:1993mu, Stephens:1993an}. 
 However, the present calculation will modify black hole
complementarity since the freely falling observer can perceive radiation 
throughout free fall especially at the event horizon. It means
that when the observer passes through the horizon, he/she sees that
the infalling particle may be either excited or annihilated by the 
energy density around the black hole.
 
%\section*{Acknowledgments}
 \acknowledgments %
 WK was indebted to Shailesh Kukarni, Andrei Smilga, Donghan Yeom, and
 Sang-Heon Yi for exciting discussions.  
This work was supported by the National Research Foundation of Korea(NRF)
grant funded by the Korea government(MSIP) (2014R1A2A1A11049571).

%%%%%%%%%%%%%%%%%%%%%%%%%%%%%%%%%%%%%%%%%%%%%%%%%%%%%%%%%%%%%
%%%%%%%%%%%%%%%             References       %%%%%%%%%%%%%%%%
%%%%%%%%%%%%%%%%%%%%%%%%%%%%%%%%%%%%%%%%%%%%%%%%%%%%%%%%%%%%%


\begin{thebibliography}{99}


%\cite{Hawking:1974rv}
%\bibitem{Hawking:1974rv} 
 % S.~W.~Hawking,
  %``Black hole explosions,''
%  Nature {\bf 248}, 30 (1974).
 % %%CITATION = NATUA,248,30;%%

%\cite{Hawking:1974sw}
\bibitem{Hawking:1974sw} 
  S.~W.~Hawking,
  %``Particle Creation by Black Holes,''
  Commun.\ Math.\ Phys.\  {\bf 43}, 199 (1975)
  [Erratum-ibid.\  {\bf 46}, 206 (1976)].
  %%CITATION = CMPHA,43,199;%%

%\cite{Hawking:1976ra}
\bibitem{Hawking:1976ra}
  S.~W.~Hawking,
  %``Breakdown of predictability in gravitational collapse,''
  Phys.\ Rev.\  D {\bf 14}, 2460 (1976).
  %%CITATION = PHRVA,D14,2460;%%


%\cite{Page:1993df}
%\bibitem{Page:1993df}
%  D.~N.~Page,
%  %``Average entropy of a subsystem,''
%  Phys.\ Rev.\ Lett.\  {\bf 71}, 1291 (1993)
%  [arXiv:gr-qc/9305007].
%  %%CITATION = PRLTA,71,1291;%%

%\cite{Page:1993wv}
%\bibitem{Page:1993wv}
%  D.~N.~Page,
%  %``Information in black hole radiation,''
%  Phys.\ Rev.\ Lett.\  {\bf 71}, 3743 (1993)
%  [arXiv:hep-th/9306083].
%  %%CITATION = PRLTA,71,3743;%%

%\cite{Susskind:1993if}
\bibitem{Susskind:1993if}
  L.~Susskind, L.~Thorlacius and J.~Uglum,
  %``The stretched horizon and black hole complementarity,''
  Phys.\ Rev.\  D {\bf 48}, 3743 (1993)
  [arXiv:hep-th/9306069].
  %%CITATION = PHRVA,D48,3743;%%


%\cite{Stephens:1993an}
\bibitem{Stephens:1993an} 
  C.~R.~Stephens, G.~'t Hooft and B.~F.~Whiting,
  %``Black hole evaporation without information loss,''  
Class.\ Quant.\ Grav.\  {\bf 11}, 621 (1994)  [gr-qc/9310006].  %%CITATION = GR-QC/9310006;%%  %269 citations counted in INSPIRE as of 15 Oct 2013


%\cite{Susskind:1993mu}
\bibitem{Susskind:1993mu}
  L.~Susskind and L.~Thorlacius,
  %``Gedanken experiments involving black holes,''
  Phys.\ Rev.\  D {\bf 49}, 966 (1994)
  [arXiv:hep-th/9308100].
  %%CITATION = PHRVA,D49,966;%%
  

%\cite{Unruh:1980cg}
\bibitem{Unruh:1980cg} 
  W.~G.~Unruh,
  %``Notes on black hole evaporation,''
  Phys.\ Rev.\ D {\bf 14}, 870 (1976).
  %%CITATION = PHRVA,D14,870;%%
  %1799 citations counted in INSPIRE as of 12 Mar 2014



%\cite{Unruh:1977ga}
\bibitem{Unruh:1977ga} 
  W.~G.~Unruh,
  %``Origin of the Particles in Black Hole Evaporation,'' 
 Phys.\ Rev.\ D {\bf 15}, 365 (1977).  %%CITATION = PHRVA,D15,365;%%  %32 citations counted in INSPIRE as of 20 May 2014

%\cite{Candelas:1980zt}
\bibitem{Candelas:1980zt} 
  P.~Candelas,
  %``Vacuum Polarization in Schwarzschild Space-Time,''  
Phys.\ Rev.\ D {\bf 21}, 2185 (1980).  %%CITATION = PHRVA,D21,2185;%%  %187 citations counted in INSPIRE as of 20 May 2014



%\cite{Almheiri:2012rt}
\bibitem{Almheiri:2012rt} 
  A.~Almheiri, D.~Marolf, J.~Polchinski and J.~Sully,
  %``Black Holes: Complementarity or Firewalls?,''
  JHEP {\bf 1302}, 062 (2013)
  [arXiv:1207.3123 [hep-th]].
  %%CITATION = ARXIV:1207.3123;%%







%\cite{Bousso:2012as}
\bibitem{Bousso:2012as} 
  R.~Bousso,
  %``Complementarity Is Not Enough,''
  Phys.\ Rev.\ D {\bf 87}, 124023 (2013)
  [arXiv:1207.5192 [hep-th]].
  %%CITATION = ARXIV:1207.5192;%%

%\cite{Nomura:2012sw}
\bibitem{Nomura:2012sw} 
  Y.~Nomura, J.~Varela and S.~J.~Weinberg,
  %``Complementarity Endures: No Firewall for an Infalling Observer,''
  JHEP {\bf 1303}, 059 (2013)
  [arXiv:1207.6626 [hep-th]].
  %%CITATION = ARXIV:1207.6626;%%


%\cite{Banks:2012nn}
%\bibitem{Banks:2012nn}
%  T.~Banks and W.~Fischler,
%  %``Holographic Space-Time Does Not Predict Firewalls,''
%  arXiv:1208.4757 [hep-th].
%  %%CITATION = ARXIV:1208.4757;%%

%\cite{Ori:2012jx}
%\bibitem{Ori:2012jx}
%  A.~Ori,
%  %``Firewall or smooth horizon?,''
%  arXiv:1208.6480 [gr-qc].
%  %%CITATION = ARXIV:1208.6480;%%

%\cite{Susskind:2012uw}
%\bibitem{Susskind:2012uw} 
%  L.~Susskind,
%  %``The Transfer of Entanglement: The Case for Firewalls,''
%  arXiv:1210.2098 [hep-th].
%  %%CITATION = ARXIV:1210.2098;%%

%\cite{Hossenfelder:2012mr}
%\bibitem{Hossenfelder:2012mr}
%  S.~Hossenfelder,
%  %``Comment on the black hole firewall,''
%  arXiv:1210.5317 [gr-qc].
%  %%CITATION = ARXIV:1210.5317;%%

%%\cite{Nomura:2012cx}
%\bibitem{Nomura:2012cx} 
%  Y.~Nomura, J.~Varela and S.~J.~Weinberg,
%  %``Black Holes, Information, and Hilbert Space for Quantum Gravity,''
%  Phys.\ Rev.\ D {\bf 87}, 084050 (2013)
%  [arXiv:1210.6348 [hep-th]].
%  %%CITATION = ARXIV:1210.6348;%%

%\cite{Page:2012zc}
\bibitem{Page:2012zc} 
  D.~N.~Page,
  %``Hyper-Entropic Gravitational Fireballs (Grireballs) with Firewalls,''
  JCAP {\bf 1304}, 037 (2013)
  [arXiv:1211.6734 [hep-th]].
  %%CITATION = ARXIV:1211.6734;%%

%\cite{Giddings:2012gc}
%\bibitem{Giddings:2012gc} 
%  S.~B.~Giddings,
%  %``Nonviolent nonlocality,''
%  arXiv:1211.7070 [hep-th].
%  %%CITATION = ARXIV:1211.7070;%%

%\cite{Jacobson:2012gh}
%\bibitem{Jacobson:2012gh} 
%  T.~Jacobson,
%  %``Boundary unitarity without firewalls,''
 % arXiv:1212.6944 [hep-th].
 % %%CITATION = ARXIV:1212.6944;%%

%\cite{Kim:2013fv}
\bibitem{Kim:2013fv} 
  W.~Kim, B.-H.~Lee and D.-H.~Yeom,
  %``Black hole complementarity and firewall in two dimensions,''
  JHEP {\bf 1305}, 060 (2013)
  [arXiv:1301.5138 [gr-qc]].
  %%CITATION = ARXIV:1301.5138;%%

%\cite{Giddings:2013kcj}
\bibitem{Giddings:2013kcj} 
  S.~B.~Giddings,
  %``Nonviolent information transfer from black holes: a field theory parameterization,''
  Phys.\ Rev.\ D {\bf 88}, 024018 (2013)
  [arXiv:1302.2613 [hep-th]].
  %%CITATION = ARXIV:1302.2613;%%

%\cite{Almheiri:2013hfa}
\bibitem{Almheiri:2013hfa} 
  A.~Almheiri, D.~Marolf, J.~Polchinski, D.~Stanford and J.~Sully,
  %``An Apologia for Firewalls,''
  JHEP {\bf 1309}, 018 (2013)
  [arXiv:1304.6483 [hep-th]].
  %%CITATION = ARXIV:1304.6483;%%

%\cite{Verlinde:2013uja}
%\bibitem{Verlinde:2013uja} 
%  E.~Verlinde and H.~Verlinde,
%  %``Passing through the Firewall,''
%  arXiv:1306.0515 [hep-th].
%  %%CITATION = ARXIV:1306.0515;%%

%\cite{Maldacena:2013xja}
%\bibitem{Maldacena:2013xja} 
%  J.~Maldacena and L.~Susskind,
%  %``Cool horizons for entangled black holes,''
%  arXiv:1306.0533 [hep-th].
%  %%CITATION = ARXIV:1306.0533;%%

%\cite{Almheiri:2013wka}
%\bibitem{Almheiri:2013wka} 
%  A.~Almheiri and J.~Sully,
%  %``An Uneventful Horizon in Two Dimensions,''
%  arXiv:1307.8149 [hep-th].
%  %%CITATION = ARXIV:1307.8149;%%

%\cite{Giddings:2013vda}
%\bibitem{Giddings:2013vda} 
%  S.~B.~Giddings,
%  %``Statistical physics of black holes as quantum-mechanical systems,''
%  arXiv:1308.3488 [hep-th].
%  %%CITATION = ARXIV:1308.3488;%%

%\cite{Braunstein:2013bra}
\bibitem{Braunstein:2013bra} 
  S.~L.~Braunstein, S.~Pirandola and K.~Zyczkowski,
  %``Better Late than Never: Information Retrieval from Black Holes,''
  Phys.\ Rev.\ Lett.\  {\bf 110}, 101301 (2013).
  %%CITATION = PRLTA,110,101301;%%









%\cite{Hutchinson:2013kka}
\bibitem{Hutchinson:2013kka} 
  J.~Hutchinson and D.~Stojkovic,
  %``Icezones instead of firewalls: extended entanglement beyond the event horizon and unitary evaporation of a black hole,''
  arXiv:1307.5861 [hep-th].
  %%CITATION = ARXIV:1307.5861;%%
  %2 citations counted in INSPIRE as of 26 Oct 2014

%\cite{Page:2013mqa}
\bibitem{Page:2013mqa} 
  D.~N.~Page,
  %``Excluding Black Hole Firewalls with Extreme Cosmic Censorship,''
  JCAP {\bf 1406}, 051 (2014)
  [arXiv:1306.0562 [hep-th]].
  %%CITATION = ARXIV:1306.0562;%%
  %24 citations counted in INSPIRE as of 26 Oct 2014



%\cite{Brynjolfsson:2008uc}
\bibitem{Brynjolfsson:2008uc} 
  E.~J.~Brynjolfsson and L.~Thorlacius,
  %``Taking the Temperature of a Black Hole,''  
JHEP {\bf 0809}, 066 (2008)  [arXiv:0805.1876 [hep-th]].  %%CITATION = ARXIV:0805.1876;%%  %11 citations counted in INSPIRE as of 06 Nov 2013


%\cite{Greenwood:2008zg}
\bibitem{Greenwood:2008zg}
  E.~Greenwood and D.~Stojkovic,
  %``Hawking radiation as seen by an infalling observer,''
  JHEP {\bf 0909}, 058 (2009)
  [arXiv:0806.0628 [gr-qc]].
  %%CITATION = ARXIV:0806.0628;%%

%\cite{Barbado:2011dx}
\bibitem{Barbado:2011dx} 
  L.~C.~Barbado, C.~Barcelo and L.~J.~Garay,
  %``Hawking radiation as perceived by different observers,''  
Class.\ Quant.\ Grav.\  {\bf 28}, 125021 (2011)  [arXiv:1101.4382 [gr-qc]].  %%CITATION = ARXIV:1101.4382;%%  %9 citations counted in INSPIRE as of 06 Nov 2013
%\cite{Callan:1992rs}

%\cite{Smerlak:2013cha}
\bibitem{Smerlak:2013cha} 
  M.~Smerlak,
  %``The two faces of Hawking radiation,''
  Int.\ J.\ Mod.\ Phys.\ D {\bf 22}, 1342019 (2013)
  [arXiv:1307.2227 [gr-qc]].
  %%CITATION = ARXIV:1307.2227;%%
  %1 citations counted in INSPIRE as of 04 Apr 2014


%\cite{Smerlak:2013sga}
\bibitem{Smerlak:2013sga} 
  M.~Smerlak and S.~Singh,
  %``New perspectives on Hawking radiation,''
  Phys.\ Rev.\ D {\bf 88}, 104023 (2013)
  [arXiv:1304.2858 [gr-qc]].
  %%CITATION = ARXIV:1304.2858;%%
  %4 citations counted in INSPIRE as of 04 Apr 2014


%\cite{Kim:2013caa}
\bibitem{Kim:2013caa} 
  W.~Kim and E.~J.~Son,
  %``Freely Falling Observer and Black Hole Radiation,'' 
 Mod.\ Phys.\ Lett.\ A {\bf 29}, 1450052 (2014)  [arXiv:1310.1458 [hep-th]].  %%CITATION = ARXIV:1310.1458;%%  %3 citations counted in INSPIRE as of 22 May 2014


\bibitem{Callan:1992rs}
  C.~G.~Callan, Jr., S.~B.~Giddings, J.~A.~Harvey and A.~Strominger,
  %``Evanescent black holes,''
  Phys.\ Rev.\ D {\bf 45}, 1005 (1992)
  [hep-th/9111056].
  %%CITATION = HEP-TH/9111056;%%

%\cite{Birrell:1982ix}
\bibitem{Birrell:1982ix} 
 N.D. Birrell, P.C.W. Davies, Quantum Fields In Curved Space, Cambridge Univ.
Press, Cambridge, UK, 1982.
  
  
%\cite{Christensen:1977jc}
\bibitem{Christensen:1977jc} 
  S.~M.~Christensen and S.~A.~Fulling,
  %``Trace Anomalies and the Hawking Effect,'' 
   Phys.\ Rev.\ D {\bf 15}, 2088 (1977).  %%CITATION = PHRVA,D15,2088;%%  %383 citations counted in INSPIRE as of 24 Dec 2013
 
   
     %\cite{Ford:1993bw}
\bibitem{Ford:1993bw} 
  L.~H.~Ford and T.~A.~Roman,
  %``Motion of inertial observers through negative energy,''
  Phys.\ Rev.\ D {\bf 48}, 776 (1993)
  [gr-qc/9303038].
  %%CITATION = GR-QC/9303038;%%
  %29 citations counted in INSPIRE as of 25 Nov 2013

\bibitem{pi}
 E. Poisson and W. Israel, Phys. Rev. D {\bf 415}, 1796 (1990). 





%\cite{Wipf:1998ss}
%\bibitem{Wipf:1998ss} 
%  A.~Wipf,
  %``Quantum fields near black holes,''
%  Lect.\ Notes Phys.\  {\bf 514}, 385 (1998)
 % [hep-th/9801025].
  %%CITATION = HEP-TH/9801025;%%
  %11 citations counted in INSPIRE as of 25 Nov 2013






%\cite{Ashtekar:2008jd}
%\bibitem{Ashtekar:2008jd} 
%  A.~Ashtekar, V.~Taveras and M.~Varadarajan,
%  %``Information is Not Lost in the Evaporation of 2-dimensional Black Holes,'' 
%  Phys.\ Rev.\ Lett.\  {\bf 100}, 211302 (2008)
% [arXiv:0801.1811 [gr-qc]].  
%  %%CITATION = ARXIV:0801.1811;%%

%\cite{Bilal:1992kv}
%\bibitem{Bilal:1992kv} 
%  A.~Bilal and C.~G.~Callan, Jr.,
%  %``Liouville models of black hole evaporation,''
%  Nucl.\ Phys.\ B {\bf 394}, 73 (1993)
 % [hep-th/9205089].
%  %%CITATION = HEP-TH/9205089;%%

%\cite{Kim:1995wr}
%\bibitem{Kim:1995wr} 
%  W.~T.~Kim and J.~Lee,
%  %``Hawking radiation and energy conservation in an evaporating black hole,''
%  Phys.\ Rev.\ D {\bf 52}, 2232 (1995)
%  [hep-th/9502115].
%  %%CITATION = HEP-TH/9502115;%%
%\cite{Kim:2013caa}






%%%%%%%%%%%%%%%%%%%%%%%%%%%%%%%

\end{thebibliography}
\end{document}